# Observation of the supersolid stripe phase in spin-orbit coupled Bose-Einstein condensates

Jun-Ru Li[*][§], Jeongwon Lee[*], Wujie Huang, Sean Burchesky, Boris Shteynas, Furkan Çağrı Top, Alan O. Jamison and Wolfgang Ketterle

Department of Physics, MIT-Harvard Center for Ultracold Atoms, and Research Laboratory of Electronics, MIT, Cambridge, Massachusetts 02139, USA

**Supersolidity is an intriguing concept. It combines the property of superfluid flow with the long-range spatial periodicity of solids[1], two properties which are often mutually exclusive. The original discussion of quantum crystals[2] and supersolidity focused on solid helium-4 where it was predicted that vacancies could form dilute weakly interacting Bose-Einstein condensates[1, 3]. In this system, observation of supersolidity has been elusive[†]. The concept of supersolidity was then generalized to include other superfluid systems which break the translational symmetry of space. One such system is a Bose-Einstein condensate with spin-orbit coupling, which has a supersolid stripe phase[5-8]. Despite several recent studies of this system[9-11], which studied the miscibility of the spin components[5], the presence of stripes has not been detected. Here we report the direct observation of the predicted density modulation of the stripe phase using Bragg reflection. Our work establishes a system with**

---

[*] These authors contributed equally to this work.

[§] junruli@mit.edu

[†] Although the original observation of supersolidity[3] turned out to be caused by unusual elastic properties, experiments revealed quantum plasticity and mass supertransport, probably created by superfluid flow through the cores of interconnected dislocations[1, 4]

**unique continuous symmetry breaking properties and associated Goldstone mode and superfluid behaviour.**

Supersolids are defined as systems which spontaneously break two continuous, U(1), symmetries: the internal gauge symmetry by choosing the phase of the superfluid, and the translational symmetry of space by forming a density wave. With ultracold atoms, idealized Hamiltonians can be experimentally realized and studied. Starting from superfluid Bose-Einstein condensates (BECs), several forms of supersolid have been predicted by adding interactions in the form of dipolar interactions[12,13], Rydberg interactions[14], superradiant Rayleigh scattering[15], nearest-neighbour interaction in lattices[16] and spin-orbit interactions[5-8]. Several of these proposals lead to solidity along a single spatial direction maintaining their gaseous or liquid-like properties along the other directions[i]. Supersolidity in BEC is an unusual form of matter as it combines gaseous, superfluid, and solid behaviours.

In a spin-orbit coupled Bose-Einstein condensate, the supersolid stripe phase emerges naturally in a description where spin-orbit coupling acts as a spin flip process with momentum transfer, as shown in Fig. 1a. In solid-state materials, an electron moving at velocity **v** through an electric field **E** experiences a Zeeman energy term $-\mu_B \boldsymbol{\sigma} \cdot (\mathbf{v} \times \mathbf{E})$ due to the relativistic transformation of electromagnetic fields. The Zeeman term can be written as $\alpha_{ij} v_j \sigma_i /4$ where the strength of the coupling $\alpha$ has the units of momentum. The $v_x \sigma_z$ term together with a transverse magnetic Zeeman term $\beta \sigma_x$ leads to the Hamiltonian $H = ((p_x + \alpha\sigma_z)^2 + p_y^2 + p_z^2) / 2m + \beta \sigma_x$. A unitary transformation can shift the momenta by $\alpha\sigma_z$ resulting in

$$H = \frac{P^2}{2m} + \begin{pmatrix} 0 & \beta e^{2i\alpha x} \\ \beta e^{-2i\alpha x} & 0 \end{pmatrix} \quad (1)$$

where the second term represents a spin flip process with a momentum transfer of $2\alpha$. Therefore, spin-orbit coupling is equivalent to a spin flip process with momentum transfer, which can be directly implemented for ultracold atoms using a two-photon Raman transition between the two spin states[9, 17].

---

[i] It is analogous to liquid crystals which are solids, in the symmetry-breaking sense, only with regard to one degree of freedom.

A Bose-Einstein condensate with equal populations in the two spin states shows no spatial interference due to the orthogonality of the two spin states. With spin-orbit coupling, each spin component has now two momentum components (0 and either $+2\alpha$ or $-2\alpha$, where the sign depends on the initial spin state), which form a stationary spatial interference pattern with a wavevector of $2\alpha$ (Fig.1a). Such spatial periodicity of the atomic density can be directly probed with Bragg scattering[18] as shown in Fig. 1b. The position of the stripes is determined by the relative phase of the two condensates. This spontaneous phase breaks continuous translational symmetry. The two broken U(1) symmetries are reflected in two Goldstone modes, one for density (or charge), the other one for spin transport[19]. Adding a longitudinal Zeeman term $\delta_0 \sigma_z$ to (1) leads to a rich phase diagram as a function of $\delta_0$ and $\beta$[6,20]. For sufficiently large $|\delta_0|$, the ground state is in a plane-wave phase. This phase has a roton gap[10, 19], which decreases when $|\delta_0|$ is reduced, causing a roton instability and leading to a phase transition into the stripe phase.

Most experimental studies of spin-orbit coupling with ultracold atoms used two hyperfine ground states coupled by a two-photon Raman spin flip process[9-11, 21-24]. All previous studies with bosons used $^{87}$Rb[9-11]. Alkali atoms, especially those with small fine structure splitting, suffer from heating due to spontaneous emission by the near resonant laser beams. Another possible limitation for observing the stripe phase is the required miscibility of the two spin components. The difference in energy density between a BEC in the stripe phase and a phase-separated phase is $g\,\delta n^2 - (g - g_{\uparrow\downarrow}) n^2$ where $g = 4\pi\hbar^2 a / m$ and $g_{\uparrow\downarrow} = 4\pi\hbar^2 a_{\uparrow\downarrow} / m$ parameterize the interaction energy strengths between atoms in the same and in different spin states, respectively. Here $a$ ($a_{\uparrow\downarrow}$) is the interspin (intraspin) s-wave scattering length. The extra mean-field energy due to a modulation of the density n with amplitude $\delta n$ leads to phase separation when the contrast of the stripes $\eta = \delta n / n$ exceeds $\sqrt{(g - g_{\uparrow\downarrow})/g}$. For $^{87}$Rb in the $|F=1, m_F = 0\rangle$ and $|F=1, m_F = -1\rangle$ states, $(g - g_{\uparrow\downarrow}) / g = 10^{-3}$ is extremely small. In addition, the full width in $\delta_0$ for stable stripes is $W = 2 n (g - g_{\uparrow\downarrow})$ which is ~ 10Hz for $^{87}$Rb and requires extreme control of ambient magnetic field fluctuations. For these reasons the stripe phase has not yet been observed in previous studies of spin-orbit coupled rubidium atom[9-11].

All of these limitations were recently addressed by a new spin-orbit coupling scheme where orbital states (the lowest two eigenstates in an asymmetric double well potential) are used as the pseudospins[25]. Since the eigenstates mainly populate the different wells, their interaction strength $g_{\uparrow\downarrow}$ is small and can be adjusted by adjusting their spatial overlap, improving the miscibility. Furthermore, since both pseudospin states have the same hyperfine state, there is no sensitivity to external magnetic fields. The scheme is realized with a coherently coupled array of double wells using an optical superlattice, a periodic structure with two lattice sites per unit cell with inter-site tunnelling J, as shown in Fig. 2a. The superlattice has two low lying bands, split by the energy difference Δ between the double wells, each hosting a BEC in the respective band minima. The BECs in the lower and upper band minima are the pseudospin states in our system. Spin-orbit coupling and the supersolid stripes are created for the free-space motion in the 2D plane orthogonal to the superlattice. In this work, we do not use the direction along the superlattice as an external degree of freedom. The advantages of working with a stack of coherently coupled double wells are twofold: the increased signal to noise ratio and the use of interference between the double wells to suppress a background to the Bragg signal (see below).

The experimental setup is described in ref. 25. Approximately $1\times10^5$ Bose-Einstein condensed $^{23}$Na atoms were loaded into the optical superlattice along the z direction, equally split between the two pseudospin states with a density n ~ $1.5 \times 10^{14}$ cm$^{-3}$. The superlattice consists of laser beams at wavelengths of 1064 nm and 532 nm resulting in a lattice constant of d = 532 nm. Spin-orbit coupling was induced by two $\lambda_{IR}$ = 1064 nm Raman laser beams along the x and z axes, providing a momentum transfer $\hbar \mathbf{k}_{Raman} = \hbar$ ($k_{IR}$, 0, $k_{IR}$) and spin flip from one side of the double-well to the other with two-photon Rabi frequency Ω. Here $\hbar k_{IR} = 2\pi\hbar/\lambda_{IR}$ is the recoil momentum from a single IR photon.

The scheme realizes the spin-orbit Hamiltonian (1) with α = $k_{IR}$/2, β = $\left(1/\sqrt{2}\right) J\Omega/\Delta$, and an extra Zeeman term $\delta_0 \sigma_z$ = (δ−Δ)/2 $\sigma_z$ depending on the Raman beams detuning δ and superlattice offset Δ. The parameters J, Ω and Δ are determined from calibration experiments[25]. The main addition to ref. 25 was a separate laser beam in the x-y plane for the detection of the supersolid stripes. The stripes form perpendicularly to the superlattice with a periodicity of approximately $2d = 1064$ nm. Their detection requires near resonant yellow light ($\lambda_{Bragg} = 589$ nm) at an incident angle, $\theta = 16°$, fulfilling the Bragg condition $\lambda_{Bragg} = 4d \sin(\theta)$. A major challenge was the alignment of this beam to an accuracy of better than ~ 0.5°, the angular width of the Bragg signal, without any auxiliary density modulation at the same periodicity[ii]. A second challenge was the smallness of the detected signal, on the order of 10 photons.

Figure 1b shows the angular distribution of the Rayleigh scattered light induced by the 589 nm laser at $\delta_0=0$ going into the Bragg direction. The angular distribution is recorded by first focusing an imaging system onto the detection camera and then moving the camera out of focus. Without spin-orbit coupling, only Rayleigh scattering was observed (which has a broad dipolar angular pattern), filling the full aperture of the imaging system. The spin-orbit coupling leads to supersolid stripes and causes a specular reflection of the Bragg beam, observed as a sharp feature in the angular distribution of the Rayleigh scattered light (Fig 1b). The angular width (FWHM) of the observed peak of $9\pm1$ mrad is consistent with the diffraction limit of $\lambda_{Bragg} / D$, where D is the FWHM size of the cloud, demonstrating phase-coherence of the stripes throughout the whole cloud. This observation of the Bragg reflected beam is the main result of the paper, and constitutes a direct observation of the stripe phase with long-range order. For the same

---

[ii] Creating such a density modulation would have required a standing wave of laser light at 2128 nm.

parameters, we observe sharp momentum peaks in time-of-flight[25], the signature of Bose-Einstein condensates, implying superfluidity.

Our detection of the stripe phase is almost background free, since all other density modulations have different directions, as depicted in Fig. 2a. The superlattice is orthogonal to the stripes, along the z-axis. The Raman beams form a moving lattice and create a propagating density modulation at an angle of 45º to the superlattice in the $\hat{x}+\hat{z}$ direction. The pseudospin state in the upper band of the superlattice forms at the minimum of the band at a quasimomentum of q = π / d. The wave vector of the stripes is the sum of this quasimomentum and the momentum transfer that accompanies the spin flip of the spin-orbit coupling interaction[25], resulting in a stripe wave vector in the x direction. Since the difference in the wave vectors between the off-resonant density modulation and the stripes is not a reciprocal lattice vector, one cannot fulfil the Bragg condition simultaneously for both density modulations. In other spin-orbit coupling schemes where the two pseudospins do not have different quasimomenta, the off-resonant density modulation and the stripes would be collinear. The background free Bragg detection of the stripes depends on the realization of a coherent array of planer spin-orbit coupled systems.

For a pure condensate, the contrast of the density modulation is predicted[5,6] to be η = 2β/$E_r$, which is ~8 % for β ≈ 300 Hz. Here $E_r$ =7.6 kHz is the $^{23}$Na recoil energy for a single 1064 nm photon. A sinusoidal density modulation of η$N_{bec}$ atoms gives rise to a Bragg signal equivalent to γ (η$N_{bec}$)$^2$ / 4, where γ  is the independently measured Rayleigh scattering signal per atom per solid angle, and the factor ¼ is the Debye-Waller factor for a sinusoidal modulation. In Fig. 2b, we observed the expected behaviour of the Bragg signal to be proportional to $N_{bec}^2$ with the appropriate pre-factors. The prediction for the signal assumes that the stripes are long-range ordered throughout the whole cloud. If there were *m* domains, the signal would be *m* times smaller. Therefore, the observed strength of the Bragg signals confirms the long-range coherence already implied by the sharpness of the angular Bragg peak. Another way to quantify the Bragg signal is to define the ratio of the peak Bragg intensity to the Rayleigh intensity as "gain," which is calculated to be $N_{total} (f \beta/E_r)^2$ where $f = N_{bec}/N_{total}$ is the condensate fraction. The inset of Fig. 2b shows the normalized gain as a function of condensate fraction squared. The linear fit to the data points is consistent with a y-intercept of zero. This shows that the observed gain comes only from the superfluid component of the atomic sample.

We observed a lifetime of about 20 ms for the Bragg signal after ramping up the spin-orbit coupling, accompanied by a clearly visible reduction in the number of atoms in the BEC. We believe that it is limited by the heating due to the Raman driving[25]. At values of $\beta \approx 300$ Hz, the moving Raman lattice has a depth of ~3 $E_r$ which is comparable to the stationary lattice at ~10 $E_r$. When the spin-orbit coupling was increased further, the Bragg signal decreased, as shown in Fig. 2c, with noticeable atom loss. In addition, the observed heating may still have a contribution from technical sources, since the observed lifetime is sensitive to alignment.

Fig. 3a shows the phase diagram for spin-orbit coupled BECs for the parameters implemented in this work. It features a wide area for the stripe phase, much larger than for $^{87}$Rb, due to the high miscibility of the two orbital pseudospin states. The situation for our system and $^{87}$Rb are complementary. In $^{87}$Rb, the phase separated and the single minimum states were easily observed[9,10], whereas our scheme favours the stripe phase.

Exploring the phase diagram in the vertical direction requires varying the detuning δ of the two Raman beams. For $\delta_0 = 0$, the two dressed spin states are degenerate. For sufficiently large values of $|\delta_0|$, the ground state is the lower dressed spin state. The vertical width of the stripe phase in Fig. 3a depends on the miscibility of the two spin components[6, 20]. However, as observed in ref. 9, population relaxation between the two spin states is very slow. For our parameters, the equal population of the two pseudospin states is constant during the lifetime of the system for all detunings $|\delta_0|$ studied (up to ±10 kHz). Therefore, the detection of the stripes are possible even for large detuning.

When the Raman beams detuning δ was varied, we observed peaked Bragg reflection at $\delta_0 \approx \pm 0.7\ E_r$ which were characterized previously as spin flip resonances coupling $|\uparrow, p = 0\rangle$ to $|\downarrow, p = -\hbar k_{IR}\rangle$ and $|\downarrow, p = 0\rangle$ to $|\uparrow, p = \hbar k_{IR}\rangle$ (Fig 3b). The peaks in enhancement show that density modulations are resonantly created either in the $|\uparrow\rangle$ or $|\downarrow\rangle$ states. In addition, we observed a third peak around $\delta_0 = 0$ where the stripe pattern is stationary. For finite $\delta_0$, it moves at a velocity of $\delta_0/k_{IR}$. Our observation shows that the stationary stripe pattern is either more stable or has higher contrast[iii] compared to a moving stripe.

The periodicity of the supersolid density modulation can depend on external, single-atom, and two atom parameters. In the present case, the periodicity is given by the wavelength and geometry of the Raman beams. It is then further modified by the spin gap parameter β and the interatomic interactions[5, 6] to be $2d/\sqrt{1-(\beta/F)^2}$, where F = ($2E_r$ + n(g+

---

[iii] Since the tunnel coupling along the superlattice direction is weak (about 1 kHz) it seems possible that the alignment of moving stripe patterns is more sensitive to perturbations than for stationary stripes and leads to a reduced Debye-Waller factor for moving stripes.

$g_{\uparrow\downarrow}))/4$. For $\beta \approx 300$Hz, the correction due to the square root is only 0.4% and was not detected in our work. In the case of dipolar supersolids, the periodicity of the density modulation is determined by the strength of the two-body dipolar interactions[12, 13]. For supersolids created by superradiant Rayleigh scattering, the periodicity is half the wavelength of the scattered light, but modified by the atomic density through the index of refraction[15].

So far, we have presented a supersolid which breaks the *continuous* translational symmetry in the *free-space* x-direction (Fig. 2a). Unrelated to the presence of spin-orbit coupling, our superlattice system also breaks a *discrete* translational symmetry along the *lattice* z-direction (Fig. 4a), by forming a spatial period which is twice that of the external lattice. This fulfils the definition of a lattice supersolid[26, 27]. As discussed in our previous work[25], the two BECs in the two pseudospin states have different quasimomenta q in the superlattice: one at $q = 0$ in the lowest band, the other at $q = \pi / d$ at the Brillouin zone boundary in the upper band. The difference in quasimomenta implies an xy antiferromagnetic pseudospin texture, doubling the period of the superlattice. Due to tunnelling between the two wells, this spin texture leads to a density modulation with the same spatial periodicity (Fig. 4a) with maximum amplitude $(J/\Delta)$. This density modulation breaks the translational symmetry of the lattice (i.e. in the z-direction). With the same periodicity of 1064 nm as the supersolid stripe phase, this density modulation can be detected with the same geometry of the Bragg beam and camera, but rotated to the y–z plane. Figure 4b shows the observed enhanced light scattering due to Bragg reflection. The Bragg signal was absent right after preparing an equal mixture of the two pseudospin states, both in $q = 0$, and appeared spontaneously after the upper pseudospin state relaxed to the band minimum at q = π / d. The antiferromagnetic spin texture is rotating at frequency Δ, the energy offset between the two pseudospin states, which causes a temporal oscillation of the density modulation with spontaneous initial phase. With Bragg pulse duration shorter than 1/2Δ, the Bragg signal varied between 0 and 100 %, depending on the phase of the oscillation of the density modulation when probed. The increased fluctuation in Fig. 4b shows the random nature of the initial phase, which is consistent with spontaneous symmetry breaking.

In conclusion, we have observed the long-predicted supersolid phase of spin-orbit coupled Bose-Einstein condensates. This realizes a system that simultaneously has off-diagonal and diagonal long-range order. In the future, it will be interesting to characterize this system's collective excitations[19] and to find ways to extend it to two-dimensional spin-orbit coupling which leads to a different and rich phase diagram[28-30]. Another direction of the future research is the study of vortices and the effects of impurities and disorder in different phases of spin-orbit coupled condensates[31-34].

Note: When we presented this work at a workshop in Chicago on 9/26, we learnt about related unpublished work on supersolidity which subsequently appeared as a preprint. See Léonard, J., Morales, A., Zupancic, P., Esslinger T. & Donner, T. Supersolid formation in a quantum gas breaking continuous translational symmetry. arXiv: 1609.09053v1 (2016).

**Acknowledgments** We thank S. Stringari for helpful discussions and W. C. Burton for carefully reading the manuscript. We acknowledge the support from the NSF through the Center for Ultracold Atoms and by award 1506369, from ARO-MURI Non-equilibrium Many-body Dynamics (grant W911NF-14-1-0003) and from AFOSR-MURI Quantum Phases of Matter (grant FA9550-14-1- 0035)

**Author contributions** J. –R. Li, W. H., J. Lee, B. S., S. B., F. C. T., and A. O. J. contributed to the building of the experiment. J. –R. Li led the experimental efforts. J. Lee led the data analysis and simulations. W. H., J. –R. Li, and W. K. conceived the experiment. All authors contributed to the writing of the manuscript. J. –R. Li and J. Lee contributed equally to this work.

**Competing financial interests** The authors declare no competing financial interests.

**Author Information** Correspondence and requests for materials should be addressed to J. –R. Li. (junruli@mit.edu).

**Methods**

**Sample preparation** We prepared $N \approx 4 \times 10^5$ $^{23}$Na atoms in a crossed optical dipole trap. An equal mixture of spin up and spin down states is created using the method described in our earlier work[25]. The superlattice is adiabatically ramped up with an offset $\Delta = 0$. The final offset is then rapidly set by a fast frequency change of the infrared lattice and is determined by the intensity of the $\lambda = 1064$ nm lattice laser light and the relative phase $\phi_{SL}$ between the two lattices. The offset $\Delta$ is calibrated with the method described in ref. 25. We estimate the stability of the offset parameter $\Delta$ to be 1 kHz limited by drifts in the air pressure which affect the relative phase $\phi_{SL}$ between the two lattices. Subsequently, the Raman lasers inducing the spin-orbit coupling are adiabatically ramped up within ~10 ms followed by a variable hold time, after which the Bragg probe beam is applied.

**Bragg beam parameters and detection** The Bragg beam was chosen to be ~ 1030 MHz blue detuned from the sodium $| 3S_{1/2}, F = 1 \rangle$ to $| 3P_{3/2}, F = 2 \rangle$ transition with a linear polarization along the superlattice direction. The detuning was chosen such that the Bragg beam can propagate through the entire condensate without significant absorption or wave front distortions. The alignment of the Bragg beam required accurate prealignment by triangulation. Since the lattice supersolid is more robust than the stripe phase while having the same periodicity for the density modulation, the alignment procedure was first developed for the lattice supersolid. The same setting was rotated around the y-axis by 90 degree to probe for the stripe phase.

The Bragg reflected beam and the Rayleigh fluorescence were recorded with an Electron Multiplying CCD camera. The signal was normalized for fluctuating atom numbers using the fluorescence intensity monitored by a photomultiplier using a separate viewport. The Bragg signal was obtained by integrating the counts of the CCD pixels around the Bragg matched angle. The Rayleigh signal was obtained from fitting the diffuse background with a two dimensional Gaussian fit.

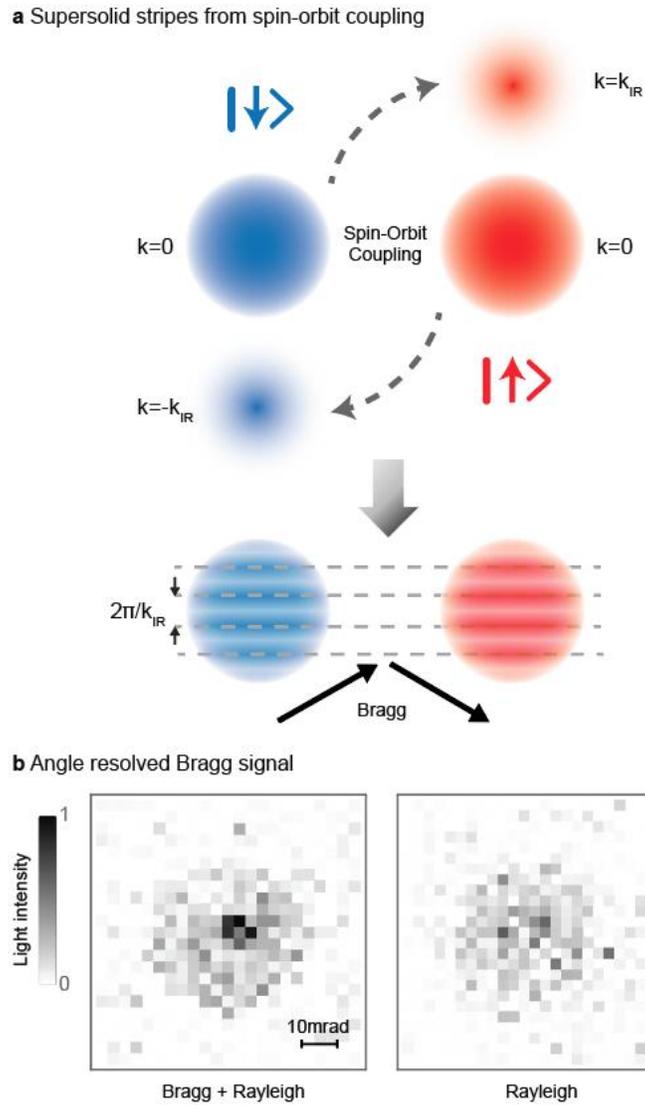

**Figure 1 | Origin of supersolid stripes and detection via Bragg scattering. a**, Spin-orbit coupling adds momentum components, $+\hbar k_{IR}$ or $-\hbar k_{IR}$ of the opposite spin state to the spin up and spin down Bose Einstein condensates (top panel: spin states in momentum space). Matter wave interference leads to a spatial density modulation of period $2\pi/k_{IR}$ (bottom panel: spin states in real space). The spatial periodicity can be directly probed by Bragg scattering. **b**, Detection of the supersolid stripe phase by angle resolved light scattering. A sharp specular feature in the left panel is the Bragg signal due to the periodic density modulation. The diffuse signal is Rayleigh scattering filling the round aperture of the imaging system. Without spin-orbit coupling, only Rayleigh scattering is observed (right panel). The figure is the average over seven shots.

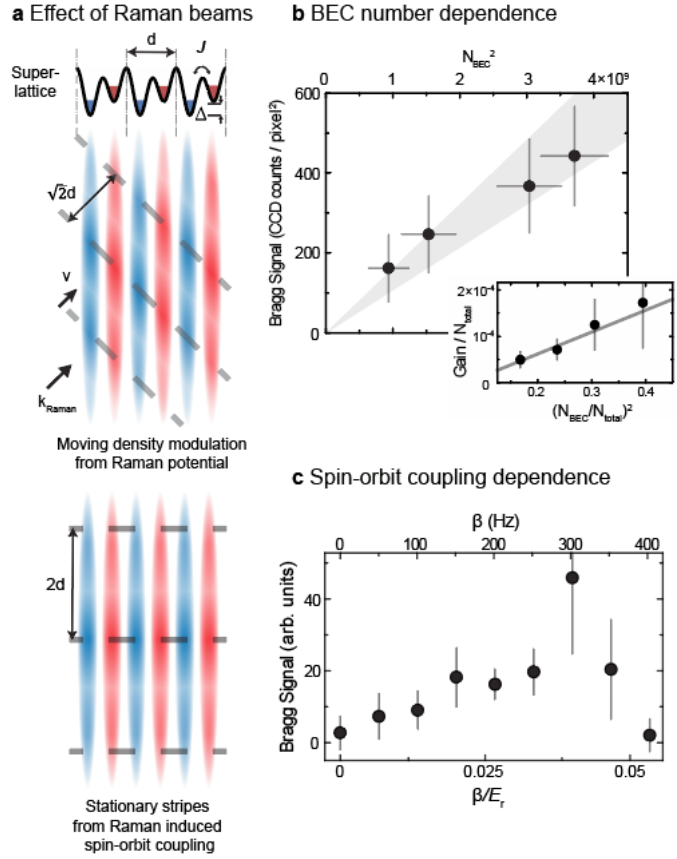

**Figure 2 | Density modulations from Raman beams, and quantitative studies of the supersolid stripes**. **a**, The two lowest bands of the superlattice are mapped into orbital pseudospins, where pseudospin down state (localized in the left wells of the superlattice) is shown in blue and pseudospin up state (localized in the right wells of the superlattice) in red. Coupling the pseudospins with Raman laser beams causes two different types of density modulations; one is a moving density modulation caused by the moving lattice potential, and the other is the stationary stripes from Raman induced spin-orbit coupling between the pseudospins. The stationary stripes along the *free-space* x-direction break the *continuous* translational symmetry. **b**, Bragg signal versus the BEC number. Shown is the count rate integrated over the Bragg peak. The grey wedge is the theoretical prediction without any adjustable parameters, using independently measured values of $\beta$ and $\gamma$ (and the corresponding $1\sigma$ errors) along with the theoretical Debye-Waller factor assuming full phase coherence of the stripes. The simple theory (see the text) predicts the peak angular amplitude of the Bragg signal. To compare it to the total count rate, we assumed a Gaussian lineshape with a constant linewidth. The linewidth was obtained by averaging the widths obtained from two-dimensional Gaussian fits to the data for each condensate number. **Inset,** Normalized gain is shown as the BEC fraction is varied. Grey solid line is a linear fit, where the y-intercept is consistent with zero within $2\sigma$ fitting error. **c**, Bragg signal versus Raman coupling strength $\beta$ at zero Raman detuning.

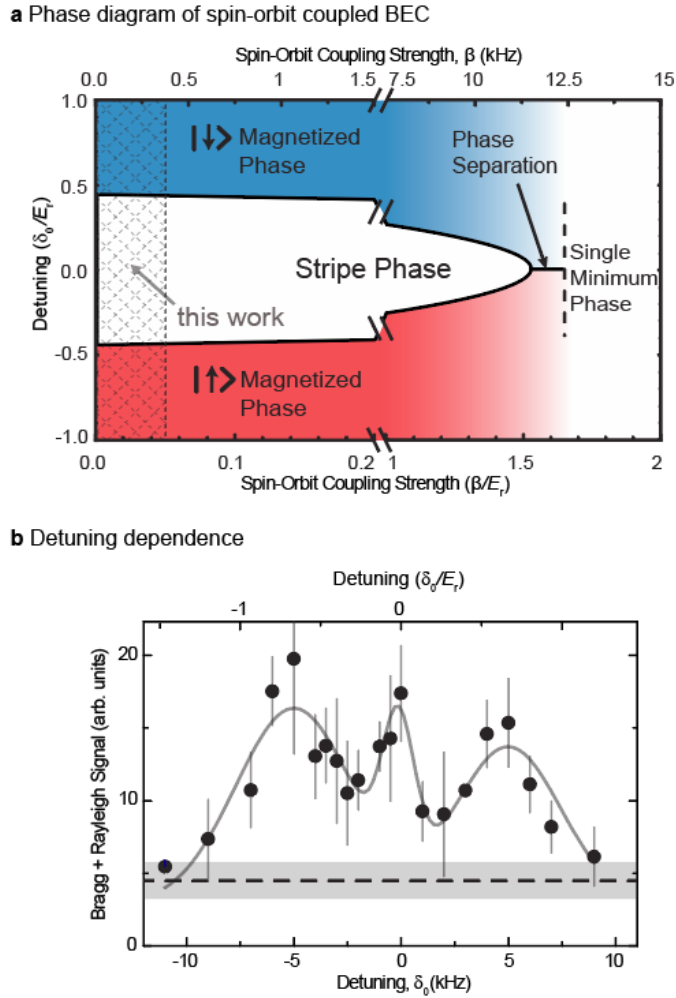

**Figure 3 | Phase diagram for spin-orbit coupled Bose-Einstein condensates, and effect of Raman detuning on the supersolid stripes**. **a**, Mean-field phase diagram of spin-orbit coupled Bose-Einstein condensates as a function of detuning and spin-orbit coupling strength. The parameter space explored in this work is shown in grey crosshatch lines. However, due to metastability, our effective detuning is always zero (see text). **b**, Black filled circles show the total light scattering signal (Bragg and Rayleigh) as a function of the frequency detuning. The light was detected within a solid angle of 10 mrad. The grey dashed line and shaded area show the mean and standard deviation of the Rayleigh scattered light for the same conditions. The light grey solid line is a triple Gaussian fit to the total light scattering where the widths and centre positions of the two outer peaks are constrained to be identical to the spin flip resonances studied in our previous work[25].

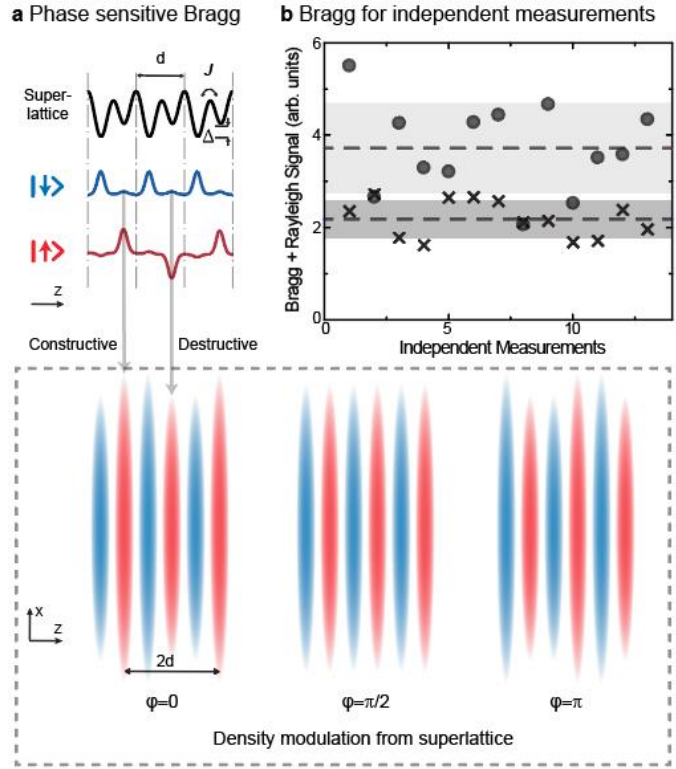

**Figure 4 | Bragg detection of a lattice supersolid caused by an antiferromagnetic spin texture**. **a**, The interference between the pseudospin down and up states alternates between constructive and destructive for adjacent unit cells in the lattice. This generates an oscillating density modulation, which has a spatial periodicity of 2d along the superlattice direction (z axis) and depends on the relative phases between the orbital pseudospins as φ =$\varphi_0$+Δ t, with $\varphi_0$ as the spontaneous phase. The density modulations at different phase conditions are shown in the grey dashed-line box. This breaks the *discrete* symmetry of the lattice potential. **b**, The Bragg signal, which is set up for detection of a spatial periodicity of 2d along the superlattice, depends on the relative phase φ when the Bragg pulse width is shorter than 1/(2Δ). Black x's show the Rayleigh scattered background integrated over 40mrad before the antiferromagnetic spin texture develops. Grey filled circles show Bragg enhanced scattering. The Bragg enhancement fluctuated between zero and a factor of two, which indicates variations of the spontaneous phase $\varphi_0$, between independent measurements.